\documentclass[preprint]{elsarticle}

\usepackage{graphicx}   
\usepackage[font=footnotesize]{subfig} 

\begin{document}
\title{Systematic uncertainties in air shower measurements from\\
  high-energy hadronic interaction models}

\author[leeds]{R.D. Parsons} 
\author[carla]{C. Bleve} 
\author[sergei1,sergei2]{S.S. Ostapchenko} 
\author[leeds]{J. Knapp} 

\address[leeds]{School of Physics and Astronomy, University of Leeds,
Leeds, LS2 9JT, UK}
\address[carla]{Dept. of Physics, Bergische Universit\"at Wuppertal, 
 42097 Wuppertal, Germany}
\address[sergei1]{Institutt for Fysikk, NTNU, 7491 Trondheim, Norway}
\address[sergei2] {D.V. Skobeltsyn Institute of Nuclear Physics, Moscow
State University, 119992 Moscow, Russia}

\date{\today}

\begin{abstract}
Hadronic interaction models at cosmic ray (CR) energies are inherently
uncertain due to the lack of a fundamental theoretical description of
soft hadronic and nuclear interactions and the large extrapolation
required from collider energies to the range of the most energetic
cosmic rays observed ($>10^{20}$ eV).  Model uncertainties
are evaluated within the QGSJET-II model, by varying some of the
crucial parameters in the limits allowed by collider data, and between
QGSJET-II and other models commonly used in air shower
simulations. The crucial parameters relate to hard processes, string
fragmentation, diffraction and baryon production.  Results on
inelastic cross sections, on secondary particle production and on the
properties of air showers measured by ground detectors from energies
of $10^{12}$ to $10^{19}$ eV are presented.
\end{abstract}

\maketitle

\section{Introduction}
 
 Direct measurements of the primary cosmic rays (CR) with energies
 $\ge 10^{12}$ eV  are generally difficult
due to their exceedingly low flux. Instead, their
properties are reconstructed from the shape and particle content of
the extensive air showers (EAS) they produce in the atmosphere.  The
reconstruction is based on numerical models of the air shower
development.  As there is currently no reliable fundamental theory
describing soft hadronic interactions at high energies, large
systematic uncertainties limit the precision of the results.
Most current hadronic interaction models use Gribov-Regge theory \cite{GRtheory}
of multi-Pomeron exchange between nucleons as the basis for the treatment
of high-energy, soft interactions, which are prevalent in air showers.
Perturbative quantum chromodynamics (pQCD) can describe hard
interactions with high p$_\perp$, which are rare in cosmic ray
interactions, but become more important at higher energies.  In
addition, diffractive interactions, collisions of nuclei, and
interactions and decays of all possible secondary particles at
energies from MeV to beyond $10^{20}$ eV are needed for a complete
simulation of an air shower.  The generalisation from nucleon-nucleon 
to hadron-nucleus and nucleus-nucleus  collisions is usually performed
via the Glauber-Gribov formalism \cite{Glauber,GRnucl}, taking into account
inelastic screening and low mass diffraction effects.  
Particle tracking and electromagnetic interactions are straight forward
to simulate.  The major problem is the seamless and coherent
combination of the different parts of hadronic models.  Different
numerical codes implement the same theoretical ideas in different
ways, with approximate agreement at lower energies where collider data
are used for tuning, but diverging in the region where extrapolations
are required, e.g. to ultra high energies, or to very forward emission
directions.

An important aspect of the analysis of experimental data,
is the estimation of systematic uncertainties 
of air shower observables, 
due to imperfections of the current interaction models. 
The sensitivity of the simulated
EAS observables to various extrapolations
of characteristics of hadronic interactions, 
such as the inelastic cross section, 
the multiplicity of secondary particles or the relative energy
loss of the leading (most energetic) hadrons,
towards very high energies has been recently investigated in 
\cite{ulrich2010}.
While that study is certainly very illustrative,
the variations considered can hardly be
achieved in realistic models, without being in 
contradiction with available accelerator data or with fundamental principles
of the quantum field theory.
For example, one of the options investigated
in  Ref.~\cite{ulrich2010} assumed the inelastic hadron-nucleus cross 
section to rise asymptotically as $\ln^3s$ ($s$ being the square of the c.m.~energy
of the collision), thus violating the unitarity bound.

In the present work a different strategy is chosen.  We investigate the
potential spread in the high energy extrapolations of hadronic
interaction characteristics and the related uncertainties of air
shower simulations within a particular model framework. 
We quantify the uncertainties of the model,
by varying some crucial model parameters
within the limits allowed by accelerator data. 
In this way, we also take
into account the correlations between different interaction characteristics. 
For example, 
constraints on the rise of the total proton-proton cross
section with energy come also from the data on 
the pseudorapidity ($\eta$) density of secondary charged hadrons.  
The main question addressed here are how uncertain predictions can be (within a
particular model) 
and if the uncertainties can cover the differences
between currently available Monte Carlo generators of hadronic interactions.
  
For our study we chose the latest version of the QGSJET model,
QGSJET-II-3 (\cite{qgsjet}, released in 2006), because QGSJET describes 
quite well a large set of experimental air shower
data at energies from 10$^{12}$ to 10$^{20}$ eV.
Additionally, QGSJET-II provides a microscopic treatment of nonlinear
interaction effects in hadronic and nuclear collisions, described by
Pomeron-Pomeron interaction diagrams, which become very important
at very high energies. We construct a number of alternative parameter
sets (options 2 to 6) for the model and compare the results obtained with 
the standard  QGSJET-II (option 1) and with two models often used in air shower
physics, namely SIBYLL 2.1 \cite{sibyll} and   EPOS 1.99 \cite{epos}. 
While the original  SIBYLL was based on the  ``minijet'' idea,
ascribing the energy evolution of the interaction cross section and 
particle production solely to the contribution of hard processes
(minijet emission), the actual model version is rather similar to 
the Gribov-Regge-type models. A crucial difference between
QGSJET-II and SIBYLL 2.1 is that parton 
saturation effects in SIBYLL are modelled by means of
an energy-dependent cutoff for the minijet production.
EPOS employs a phenomenological approach for nonlinear interaction effects,
while addressing an important problem of energy and momentum correlations
between multiple scattering processes at the amplitude level \cite{hla01}.
EPOS has a significantly larger number of
adjustable parameters than QGSJET-II,
but a substantially larger set of accelerator data has been used for its
calibration.
The distinctive feature of EPOS is the enhanced production of baryons,
which leads to more energy in the muonic component of showers \cite{pierog2008}.
The standard versions of all these models are available in the framework of
the latest CORSIKA air shower simulation package \cite{corsika}.

\section{Parameter Variations in QGSJET-II}
Among the crucial parameters responsible for the high energy
extrapolation of the model are 
those related to inelastic diffraction,
the relative contributions of soft and hard processes, and
the distribution of energy and momentum between secondary particles. 
To investigate the corresponding allowed parameter range, 
five versions (options 2-6) of the QGSJET-II model
have been created, with varying parameter settings, which are tuned to
reproduce  (within  experimental accuracy)
{\em the same set} of accelerator  data as used for the original model
calibration. The new data obtained at the Large Hadron
Collider (LHC) have not yet been used in the analysis.
The effects of the parameter variations are then investigated 
for both single interactions and air
showers as a whole.  Option 1 corresponds to the standard
parameter settings of QGSJET-II (see Tab. \ref{tab-parameters}).

\begin{table}[b]
\begin{center}
\begin{tabular}{|c|c|c|c|c|c|}
\hline
option & \multicolumn{2}{c|}{diffraction} & $Q_0^2$ &BJM& SE\\
        & \rule[-2mm]{0mm}{5mm} 
        $\frac{\lambda_1}{\lambda_2}(p)$& $\frac{\lambda_1}{\lambda_2}(\pi)$ & (GeV$^2$) & &\\
\hline
\hline
1 (std)  & 4& 4.7 & 2.5 & on & 0.5\\
\hline        
2  & {\bf 7}& {\bf 9} & 2.5 & on & 0.5\\
\hline        
3 & 4& 4.7 & {\bf 4}    &on & 0.5\\
\hline        
4  & 4& 4,7 & 2.5 & {\bf off} & 0.5\\
\hline        
5   & 4& 4.7 & 2.5 & {\bf off} & {\bf 0.7}\\
\hline        
6  & 4& {\bf 3} & 2.5 & on & 0.5\\
\hline
\end{tabular}
\end{center}
\vspace*{-.5cm}
\caption{Parameter settings of six options of QGSJET-II.
Option 1 represents the standard settings of QGSJET-II.
See text for further explanations.}
\label{tab-parameters}
\end{table}

Diffractive interactions are very important for the shower
development, as they transport the primary energy efficiently through
the atmosphere. Both, projectile and target nucleons can undergo
low-mass diffraction excitation.  The latter is treated in QGSJET-II 
in the  framework of the Good-Walker-like multi-channel eikonal approach 
\cite{goo60,kai79}: A hadron is represented by a superposition of a number
of elastic scattering eigenstates which undergo different absorption,
depending on their relative strength $\lambda_i$ for the interaction.
For the 2-component scheme realised in  QGSJET-II, the ratio 
$\lambda_1/\lambda_2$ defines thus the strength  of diffraction,
which rises for more asymmetric superposition (larger $\lambda_1/\lambda_2$)
and disappears for $\lambda_1=\lambda_2$. In addition, a stronger
diffraction implies also stronger inelastic screening effects which
reduce the predicted  total and inelastic cross sections.
The parameters are varied in two different ways to increase the
proportion of diffractive events and influence the rise with energy of
total inelastic hadron-hadron and hadron-nucleus cross sections.  In
the default version of QGSJET-II (option 1) the corresponding ratios
are $\frac{\lambda_1}{\lambda_2}(p) = 4$ for protons and
$\frac{\lambda_1}{\lambda_2} = 4.7$ and 5.7 for pions and kaons respectively. 
The corresponding low mass diffraction cross section for the incident
hadron changes between 0.9 and 1.6 mb for proton-proton and 
between 1.2 and 1.7 mb 
for pion-proton interactions in the laboratory energy range of $10^2$ to $10^8$ GeV.
In option 2 we enhance low-mass diffraction by choosing
$\frac{\lambda_1}{\lambda_2}(p) = 7$  and 
$\frac{\lambda_1}{\lambda_2}(\pi) = 9$, $\frac{\lambda_1}{\lambda_2}(K) = 12$,
while in option 6 we leave proton
diffraction unchanged but decrease diffraction for pions and kaons,
$\frac{\lambda_1}{\lambda_2}(\pi,K) = 3$.
As an illustration of the discussed modifications, in option 2
the low mass diffraction cross section of the projectile hadron 
at $10^2$ and $10^8$ GeV is increased to
1.6 and 3.1 mb, respectively, in $pp$ collisions and to 2.0 and 3.2 mb
in $\pi p$ interactions.

In the Gribov-Regge framework,
QGSJET takes into account contributions of both soft and 
semihard parton cascades. In the latter case, the virtuality ($q^2$)  
 cutoff   $Q_0^2$ defines the
transition from the non-perturbative soft ($|q|^2 <Q_0^2$)
to the  perturbative hard ($|q|^2 > Q_0^2$) part of the cascade.
Additionally, in  QGSJET-II the shadowing and saturation effects
are accounted for by the soft ($|q|^2 <Q_0^2$) partons. Thus,
the performance of the model depends
strongly on this cutoff parameter and on the parton distribution functions
obtained from deep inelastic scattering. As the semi-hard contribution
rapidly increases with energy, one can significantly reduce the 
rise of the cross sections and the secondary
particle production  with energy when this cut-off is increased from its default
value 2.5 GeV$^2$ to 4 GeV$^2$, as done in option 3.

\begin{figure}[t]
\begin{center}
\includegraphics[width=.49\textwidth]{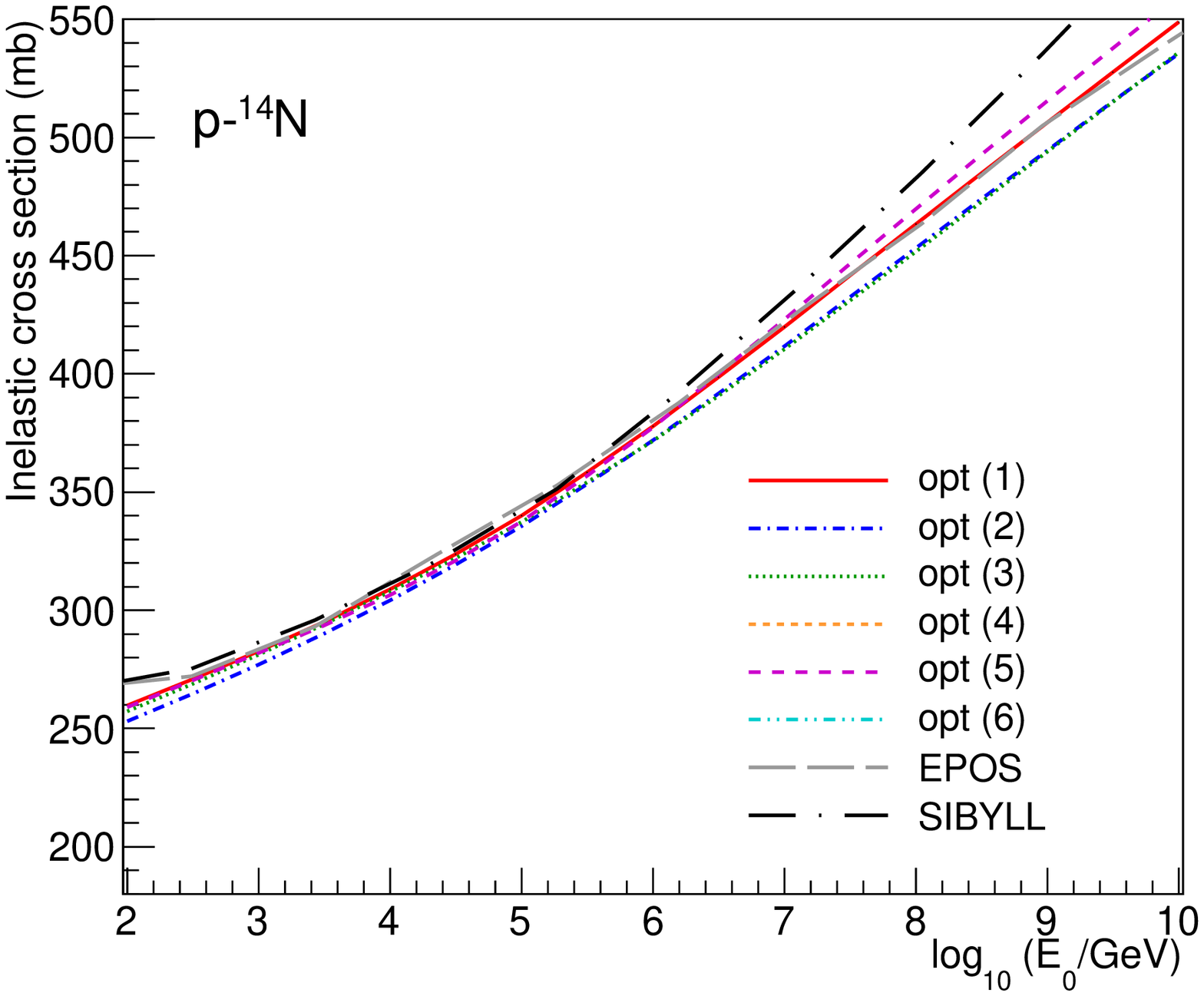} \hfill
\includegraphics[width=.49\textwidth]{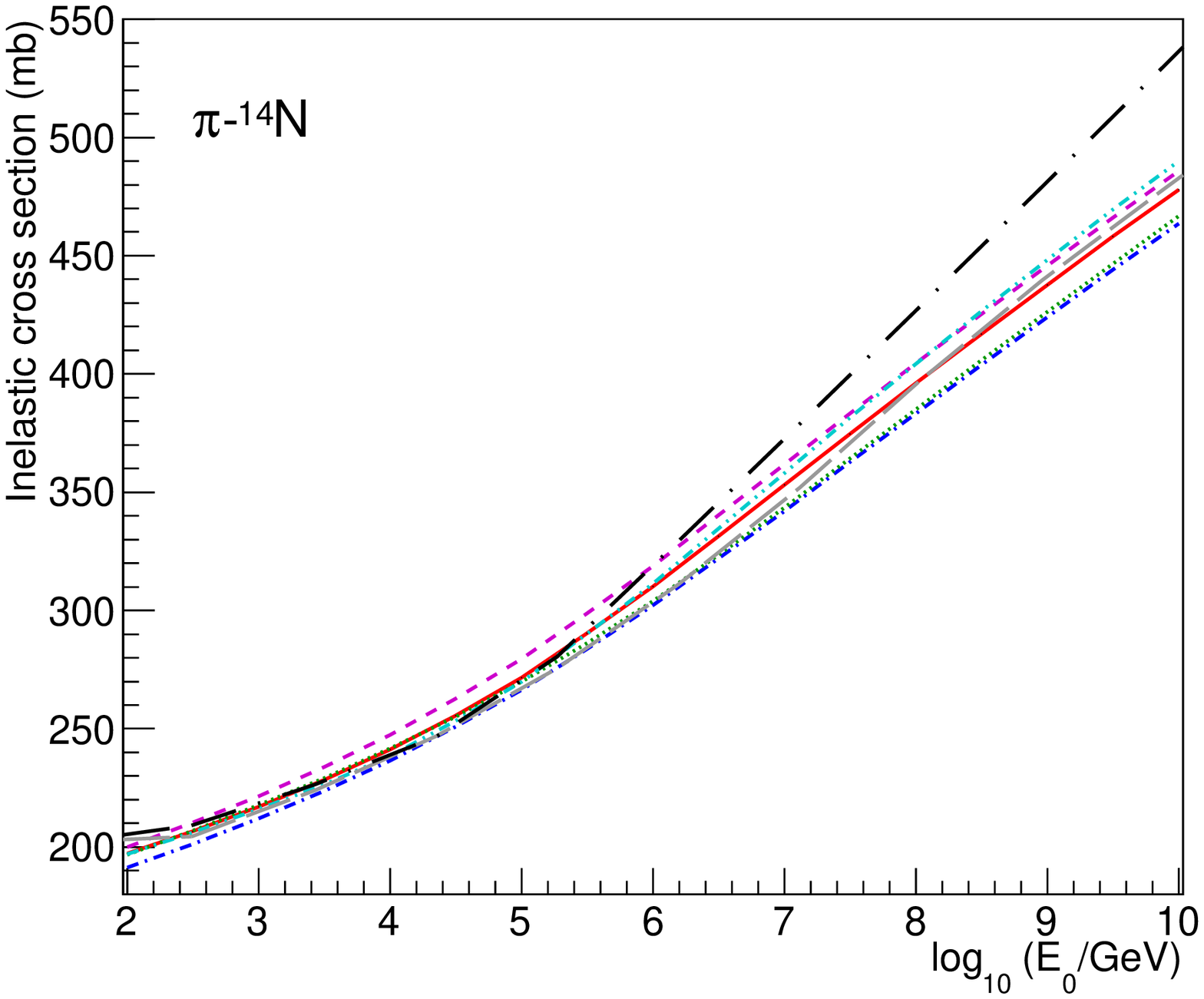} 
\includegraphics[width=.49\textwidth]{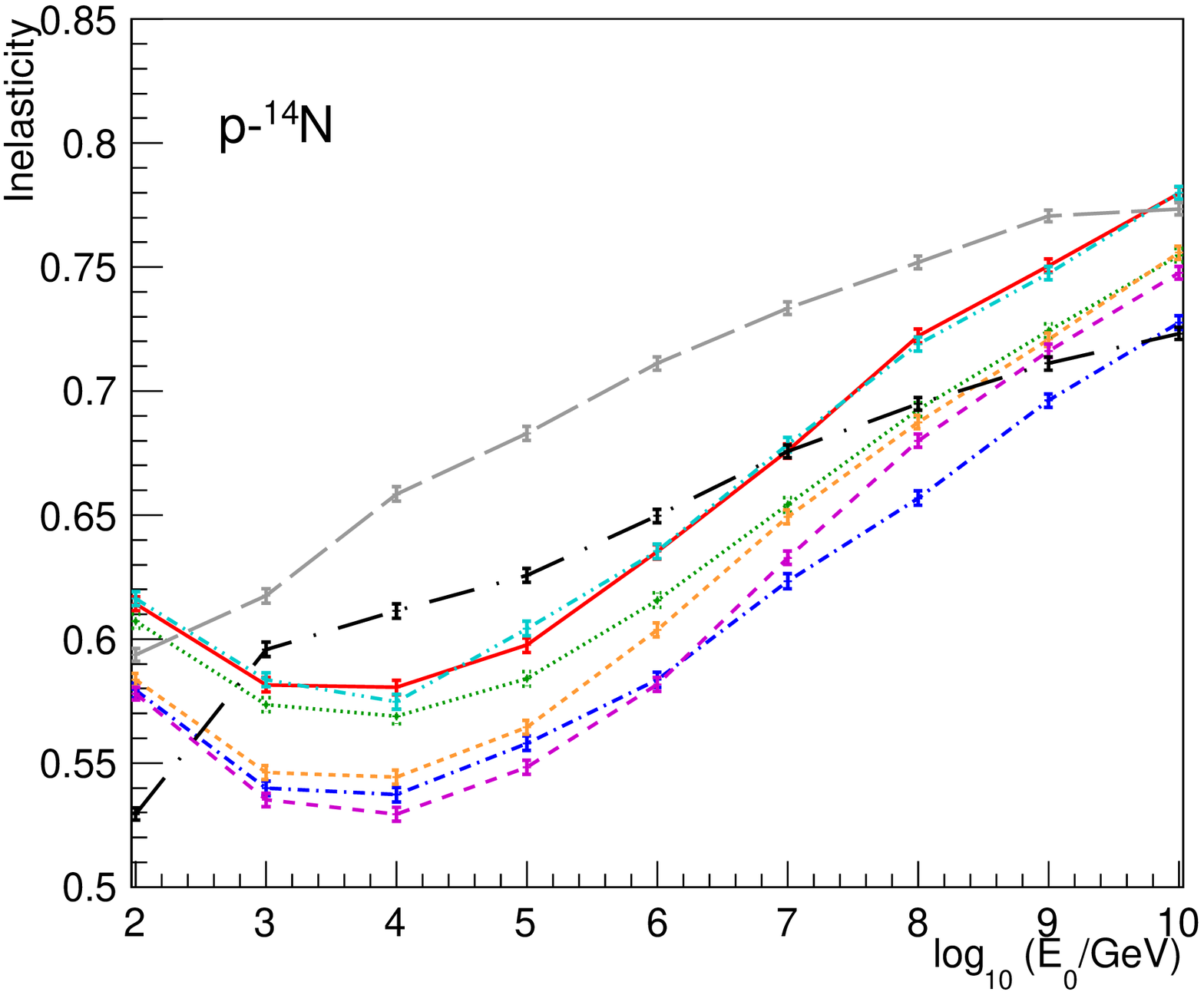} \hfill
\includegraphics[width=.49\textwidth]{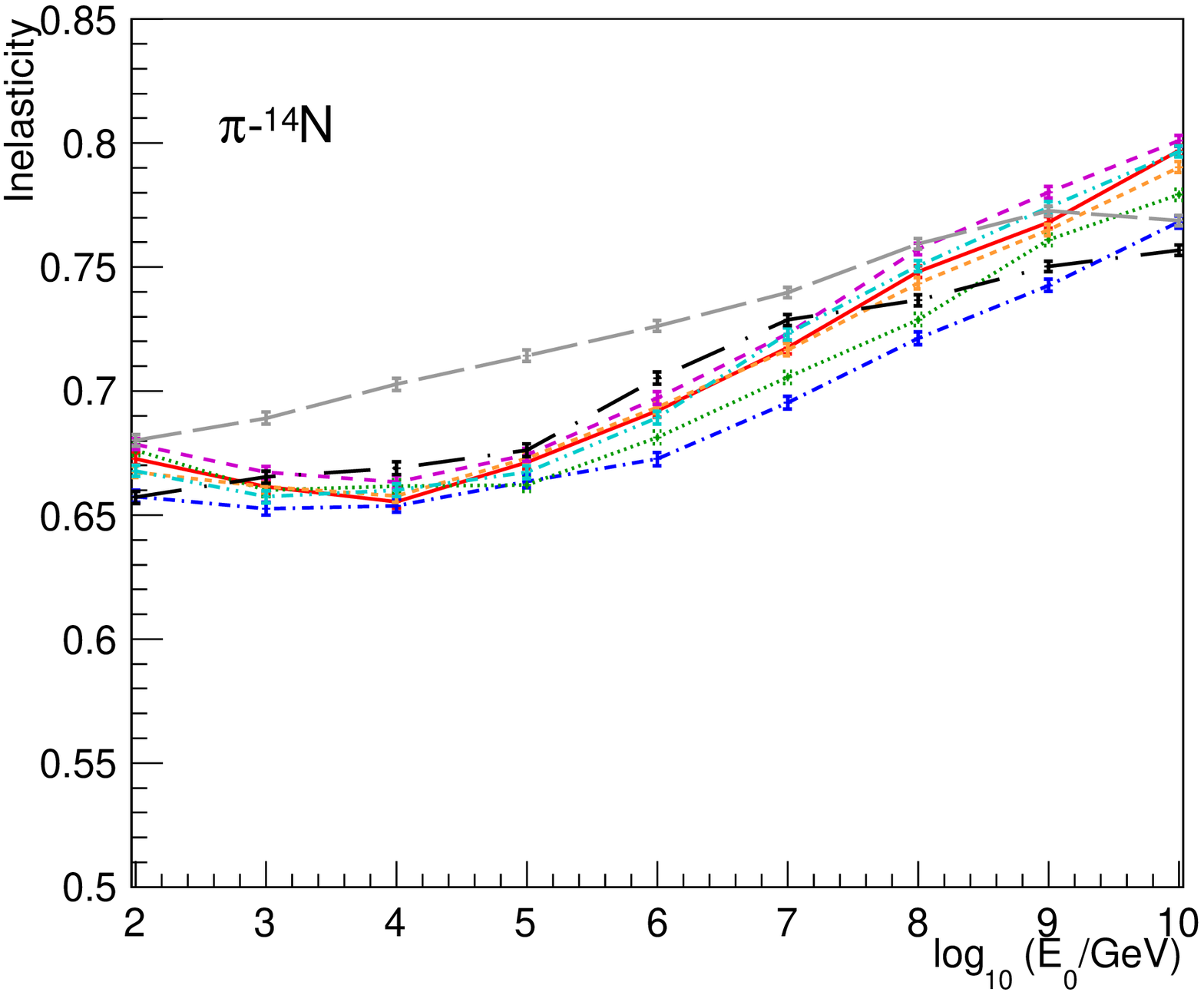} 
\end{center}
\vspace*{-0.5cm}
  \caption{\textbf{Top:} Inelastic cross sections for proton-Nitrogen 
  and $\pi$-Nitrogen collisions. \newline
 \textbf{Bottom:} Inelasticity of p-Nitrogen and $\pi$-Nitrogen collisions.}
\label{fig-cross}
\end{figure}

\begin{figure}[t]
\begin{center}
 \includegraphics[width=0.49\textwidth]{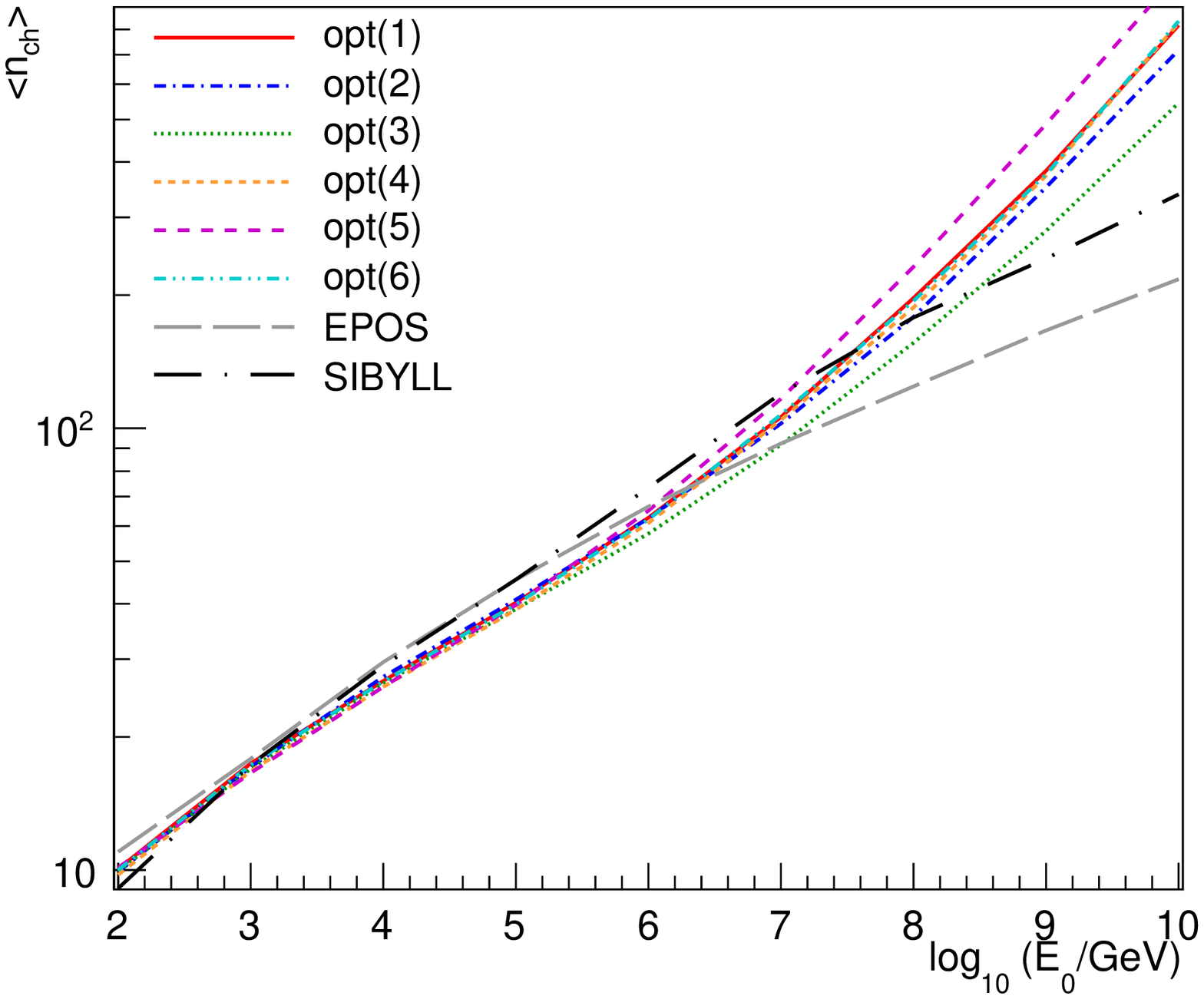} \hfill
 \includegraphics[width=0.49\textwidth]{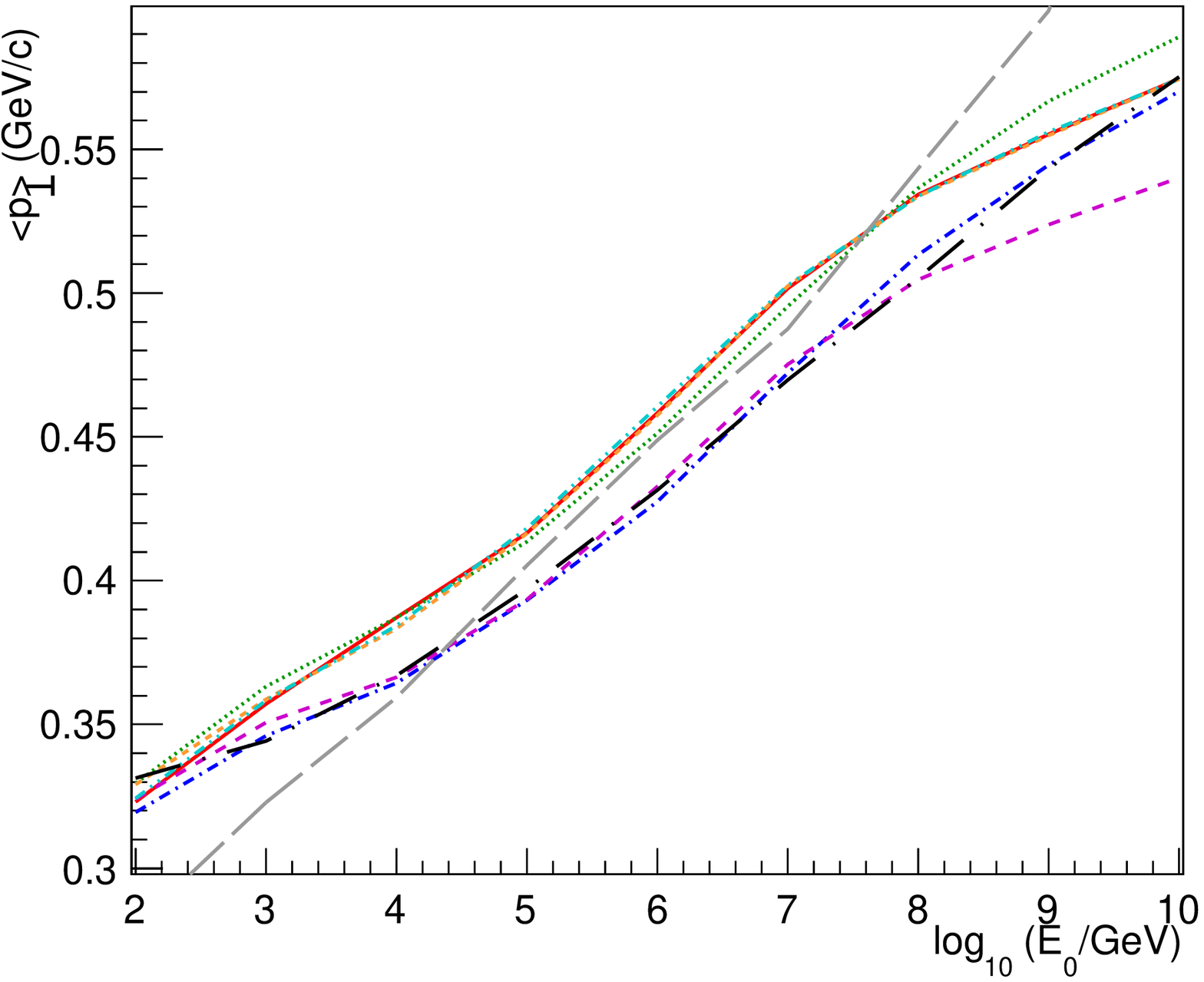}
\end{center}
\vspace*{-0.5cm}
 \caption{\textbf{Left:} Average multiplicity of charged particles and \newline
 \textbf{right:} average p$_{\perp}$ for proton-Nitrogen interactions
 as function of lab energy.}
\label{fig-nch}
\end{figure}

\begin{figure}[t]
\begin{center}
 \includegraphics[width=0.49\textwidth]{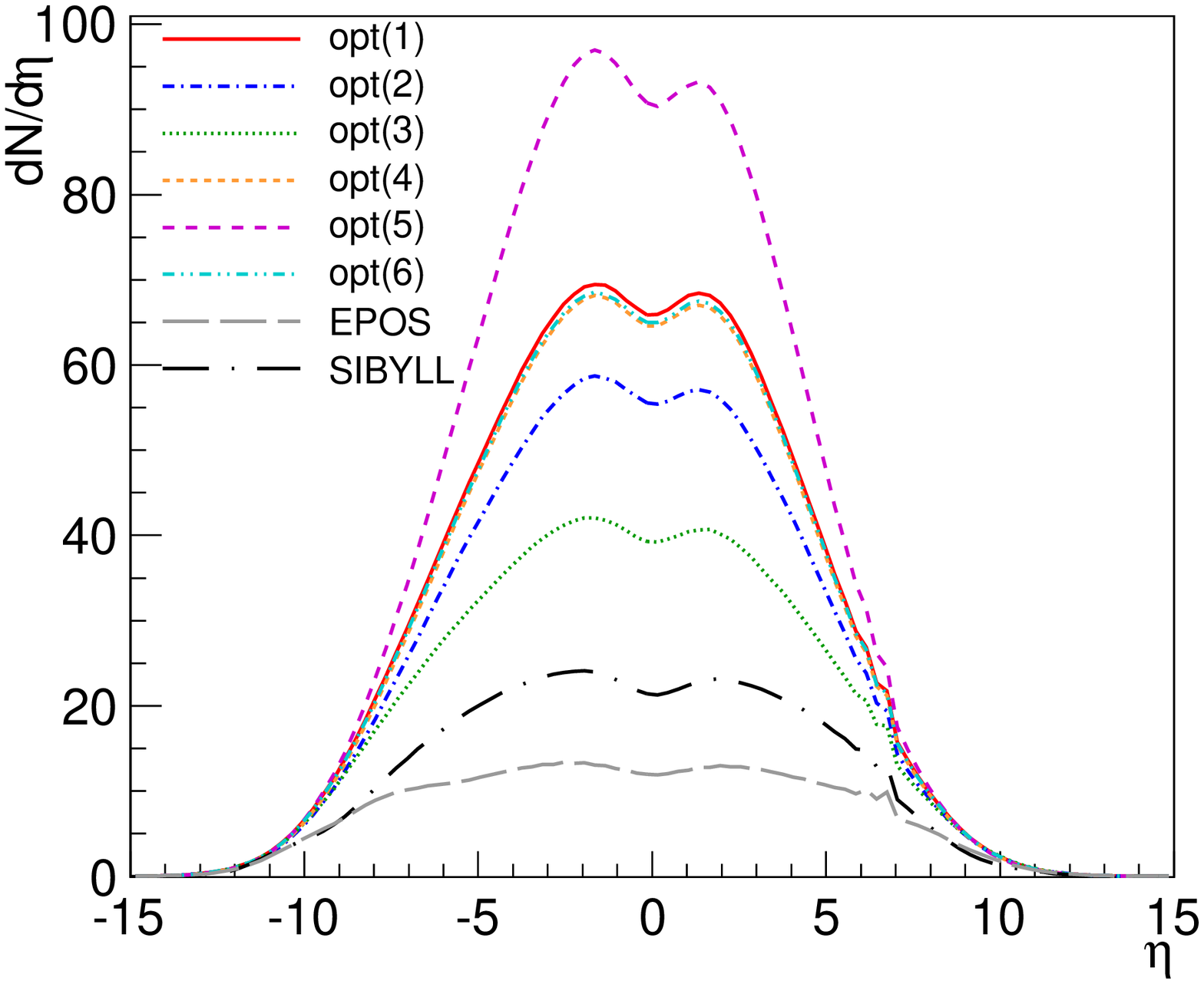} \hfill
 \includegraphics[width=0.49\textwidth]{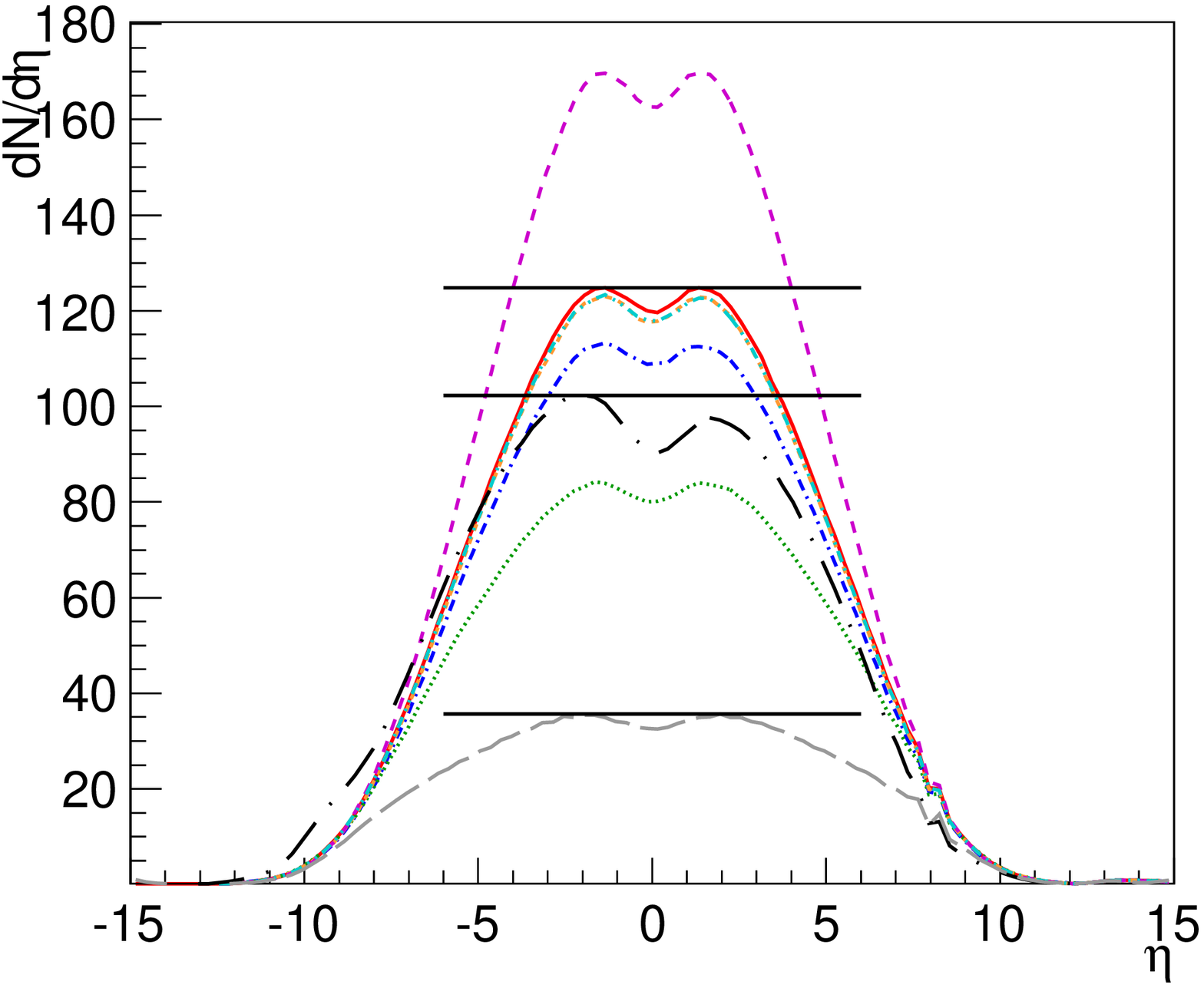} \\
 \includegraphics[width=0.49\textwidth]{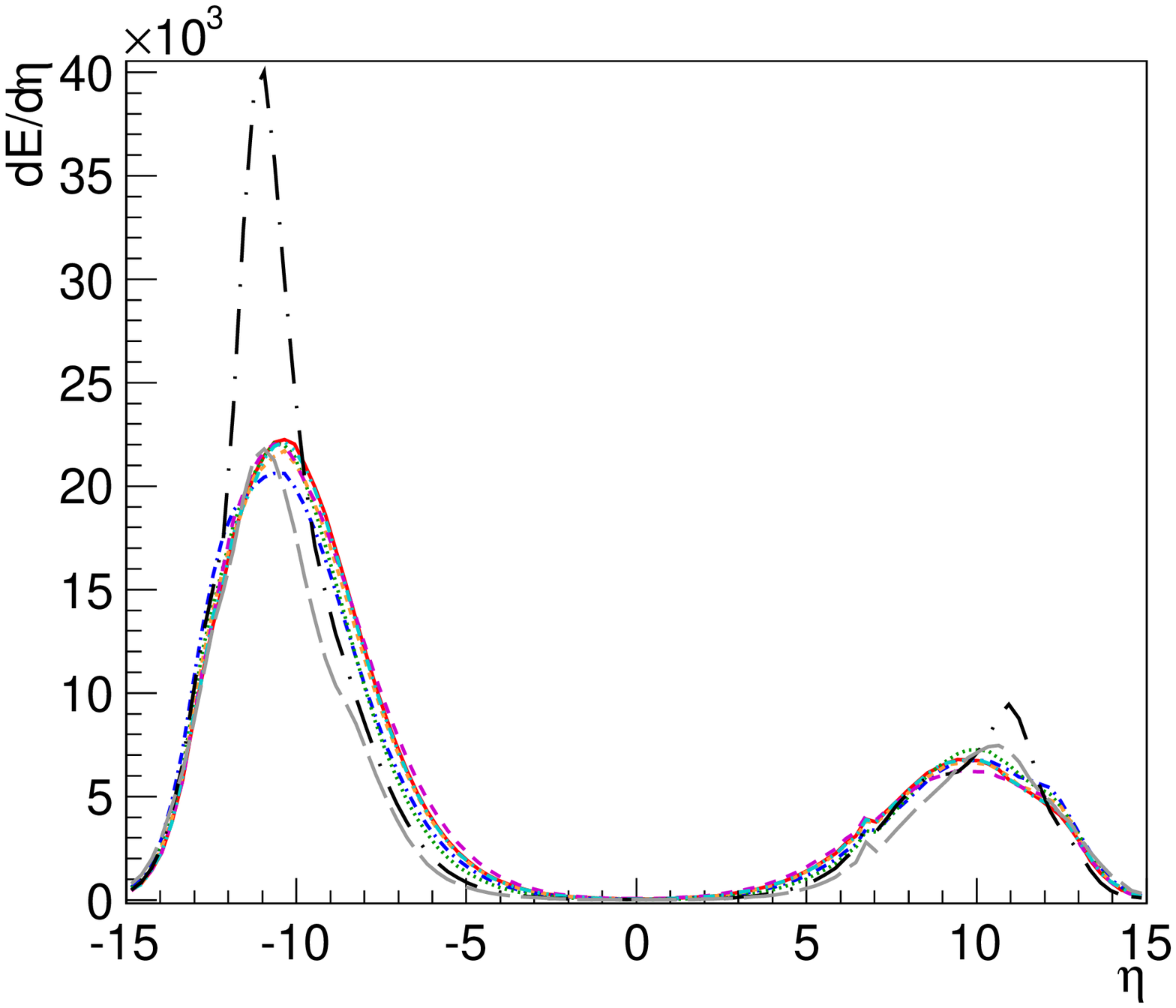} \hfill
 \includegraphics[width=0.49\textwidth]{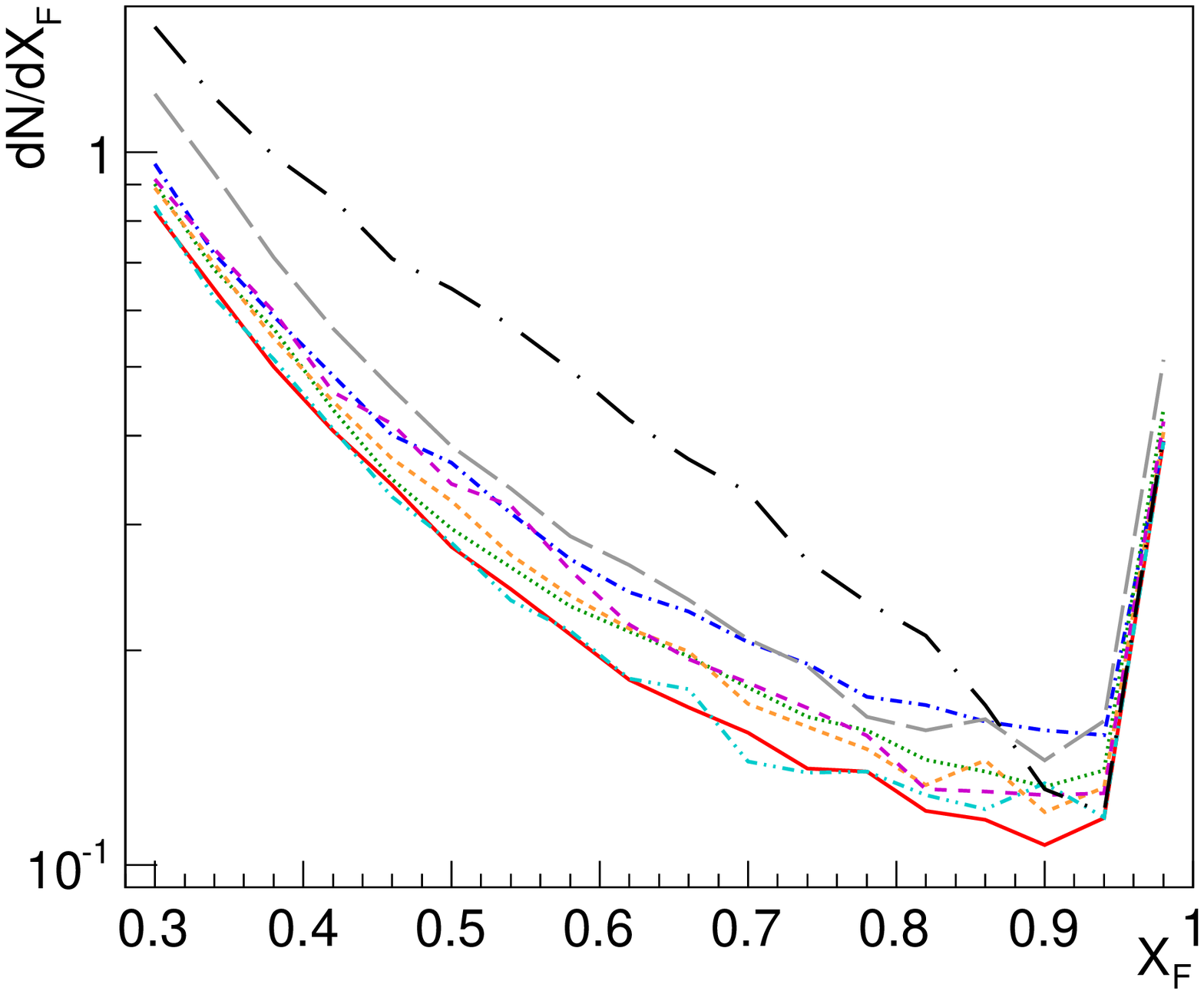}
\end{center}
 \vspace*{-0.5cm}
  \caption{\textbf{Top:}
  $dN/d\eta$  distributions 
  for charged particles from p-Nitrogen (left) and Nitrogen-Nitrogen (right) collisions at $E_{\rm lab} = 10^{19}$ eV. In these plots lines from options 1, 4 and 6 lie on top of one another. 
\textbf{Bottom:}
 $dE/d\eta$  distributions (left) and 
 Feynman-x distributions (right) 
 for charged particles from p-Nitrogen collisions. 
 All plots are for $10^{19}$ eV. Colours as for Fig. \ref{fig-nch}.}
\label{fig-ps}
\end{figure}

Options 4 and 5 vary the energy-momentum partition between elementary
production processes and string fragmentation. In option 4 the
``baryon junction'' mechanism (BJM) \cite{kha95},
related to di-quark valence quark
interactions, is switched off, leading to a slower energy rise of the
inelasticity in proton-proton and especially
in proton-nucleus collisions.
In option 5, in addition the
parameter $\alpha_q$ for the momentum distribution ($\sim x^{-\alpha_q}$)
of constituent (anti-)quarks - string
ends (SE) was modified from the default value
0.5 to 0.7, which also enhances the leading particle effect
in high energy interactions.

In addition,
with  the exception of option 4, which is identical to the default
model version apart from the above-discussed modification,
for each new parameter set created a re-tuning of other model
parameters was performed in order to keep the agreement with accelerator data.
The corresponding changes were minimal for option 6 (weaker diffraction
for pions and kaons), which concerned the Pomeron-pion interaction vertex
only. For  option 5, a significant modification of the string
fragmentation procedure was necessary to compensate the decrease of
particle multiplicity due to shorter strings. 
An opposite procedure has been used
for option 2, where the particle yield from fragmenting strings was decreased.
For options 2, 3, and 5, most of the basic model parameters,
like Pomeron intercept and slope,   Pomeron-hadron couplings, and parton
(quark and gluon) distributions in the Pomeron had to be re-tuned,
which was necessary to remain in agreement with the data on   
total and elastic hadron-proton cross sections, elastic scattering
slopes, and on the proton structure function $F_2$.

The main parameters  modified in the different
model versions are listed in Tab. \ref{tab-parameters}.
While the parameters investigated here are certainly crucial ones, it
is not clear whether they are the only ones that matter. The
comparison with EPOS shows that there are other mechanisms which can
influence significantly the shower properties and consequently affect
the interpretation of air shower experiments.

\section{Single Interactions}

In this section the particle production in individual collisions
is investigated. It relates directly to the particle content and shape of showers,
and is crucial for energy determination and composition analysis in air
shower experiments.
Overall, for all options a reasonably 
good agreement between the available collider data
at low energies and the simulations was achieved. The evolution at high energies
Fig. \ref{fig-cross} shows the inelastic cross sections and the inelasticity of
p-Nitrogen and $\pi$-Nitrogen collisions as a function of energy, for
all the models. The cross sections of the QGSJET variants
have similar shapes and the differences between 
the options  are within 10\%, even at the highest energies, with
the smallest values being obtained for options 2 and 3 (higher diffraction and
higher $Q_0$-cutoff, respectively).
EPOS shows
a cross section very similar to the QGSJET models, which could be 
expected as both models are genuinely based on Gribov-Regge theory. 
However, SIBYLL shows a rounder shape with a  much steeper increase 
at high energies, most likely being related to the adopted
parametrisation for the energy-dependent cutoff of
minijet production.

The inelasticity is the fraction of energy
used for the production of secondary particles, 
i.e. $1 - E_{\rm lp}/E_{\rm tot}$, where $E_{\rm lp}$
is the energy of the leading baryon for proton collisions and 
the leading charged meson for pion collisions. The evolution of the inelasticity as
function of $E_{\rm lab}$ is shown at the bottom of Fig. \ref{fig-cross}.
These plots show that all QGSJET model variants produce a similarly shaped dependence of 
inelasticity with energy, with a characteristic minimum around $10^3-10^4$ GeV, 
but some are shifted to lower inelasticity values by up to 0.05. 
This is also true for the inelasticity of pion-Nitrogen collisions, 
except that the shifts are much smaller.
Both, EPOS and SIBYLL, show significantly different shapes in inelasticity vs. energy,  
but with similar values as QGSJET at around 10$^{12}$ eV and 10$^{19}$ eV.

Fig. \ref{fig-nch} shows the average number of charged particles produced in 
p-Nitrogen collisions. $\langle n_{\rm ch}\rangle$ is very similar for all the models, 
up to a lab energy of  about $10^{17}$ eV, where the models start 
to diverge. At the highest energies the QGSJET variants all produce 
2-5$\times$ more secondaries than SIBYLL or EPOS. All models
predict a similar average p$_\perp$ growing with energy,
only EPOS gives smaller p$_\perp$ values below $10^4$ GeV
and larger ones above $10^8$  GeV
(see Fig. \ref{fig-nch}, right).

The large differences in particle numbers can also be seen in the pseudorapidity 
distributions of charged particles (see Fig. \ref{fig-ps}). 
QGSJET produces the largest numbers of secondaries, and there are big differences
among its variants.
At $10^{19}$ eV QGSJET 
option 5 produces almost 50\% more,
and option 3 only half
the number of particles 
than option 1
in the central region ($\eta \approx 0$).
SIBYLL and EPOS are even lower than option 3.
Recent measurements at LHC in the central pseudorapidity range 
indicate indeed that QGSJET-II tends to overestimate the central
pseudorapidity density of secondary hadrons (see \cite{david2011} for a
detailed comparison.
 However, most of these particles are emitted with low energies, compared to
 the energy of the parent hadron. Although 
differences can be also seen at larger pseudorapidities, 
these are of much smaller magnitude than in the centre. 
While being relevant for calculations of the muon content of air showers,
the hadrons in the central rapidity region are
of little importance for the longitudinal shower development.
The fact that the forward particles are more important 
can be seen from the distribution of the energy flux (dE/d$\eta$) vs.
 pseudorapidity 
in Fig. \ref{fig-ps} bottom, left. 
In the forward region, where diffractive interactions contribute and where 
most of the interaction energy is going, 
all the models are fairly close together. 
Very little energy emitted
in the central region, 
despite the large number of particles.
EPOS shows good agreement with the QGSJet variants. 
SIBYLL, however, has prominent peaks in dE/d$\eta$ in the forward and backward regions.
Also the Feynman-x distributions show that there is fair agreement
between all the models
in the very forward region ($x_{\rm F} = p_{\rm L}/p_{\rm max} \ge 0.95$).

The $\eta$ distributions also allow assessment of the treatment of nuclear 
interactions in the models, which is important
for nuclear composition studies of cosmic rays.
As Nitrogen-Nitrogen collisions are perfectly symmetric,
also the $\eta$ distribution should be symmetric.
In QGSJET nuclear interactions are modelled as two clouds of nucleons
penetrating each other and interacting. As the modeling is 
done in a symmetric way also the outcome is symmetric.
Also EPOS produces a symmetric distribution.
SIBYLL models nuclear interactions with the semi-superposition model
which treats a nuclear projectile as a superposition of individual nucleons
and a nucleus-nucleus collision as a number of independent nucleon-nucleus collisions. 
This is by construction asymmetric and therefore the Nitrogen-Nitrogen
collisions result in asymmetric pseudorapidity distributions.
SIBYLL produces $>4\times$ more particles in 
Nitrogen-Nitrogen than in p-Nitrogen collisions,
while for QGSJET and EPOS this factor is only about 2.
Thus, for composition studies based on 
differences in the muon yield in showers from different nuclei,
it is better to rely on QGSJET or EPOS.

The large differences in $\eta$ between models and the comparison with first 
collider data at high energies
indicate that some aspects of the models are not correct and need adaptation.
Although currently there are only few measurements in the relevant forward 
region  and  LHC is not yet running at its maximum energy, 
LHC results, in particular,
the results of the LHCf experiment \cite{lhcf} provide already
constraints to some of the model parameters.
Future measurements of the total proton-proton cross section and 
of particle production at $\sqrt s=14$ TeV will greatly help to tune 
the models further.

\section{Air Showers}

Showers were simulated at energies of $10^{12}$ , $10^{15}$ and
$10^{19}$ eV, i.e. at typical energies relevant to 
Cherenkov telescopes 
(such as HESS \cite{HESSexp}, VERITAS \cite{VERITASexp} and MAGIC \cite{MAGICexp}), 
small air shower arrays 
(such as KASCADE \cite{KASCADEexp} and LHAASO \cite{lhaaso}),
and experiments for the highest energies 
(e.g. the Telescope Array \cite{ta}
or the Pierre Auger Observatory \cite{Augerexp}), respectively.
The first interaction point of the primary particle was fixed
 at typical values for the
respective primary energy, to exclude the effect of shower-to-shower
fluctuations due to the varying point of first interaction and to
concentrate on differences in average shower shape. For each
energy the average longitudinal development and the lateral distribution
of particles at ground level have been determined. Also characteristic
observables were evaluated, such as the average values of the
atmospheric depth of the shower maximum, $X_{\rm max}$, the particle
number at the shower maximum, $N_{\rm max}$, the Cherenkov photon lateral 
distribution at ground level, the electron-to-muon ratio at
ground level, the energy that would be released in an Auger-like water
Cherenkov detector at 1000 m core distances ($S_{1000}$) \cite{s1000} and the time
in which the integrated signal grows from 10\% to 50\%, $t_{1/2}$ \cite{RiseTime}.
Simulations have been performed for the QGSJET variants as well as for 
SIBYLL 2.1 and EPOS 1.99 to
allow comparisons between the scale of the uncertainties due to
parameter variations within a model, and due to model-to-model
variations.

\subsection{Low Energies ($10^{12}$ eV)}

At $10^{12}$ eV very little variation is found in both the lateral
distribution and the longitudinal development,
 for the QGSJET options, with
differences being $\le$ 10\% for all quantities investigated. This is
not surprising as the energy is close to the accelerator energies at
which the models have been tuned.
\begin{figure}[b]
\begin{center}
\includegraphics[width=0.49\textwidth]{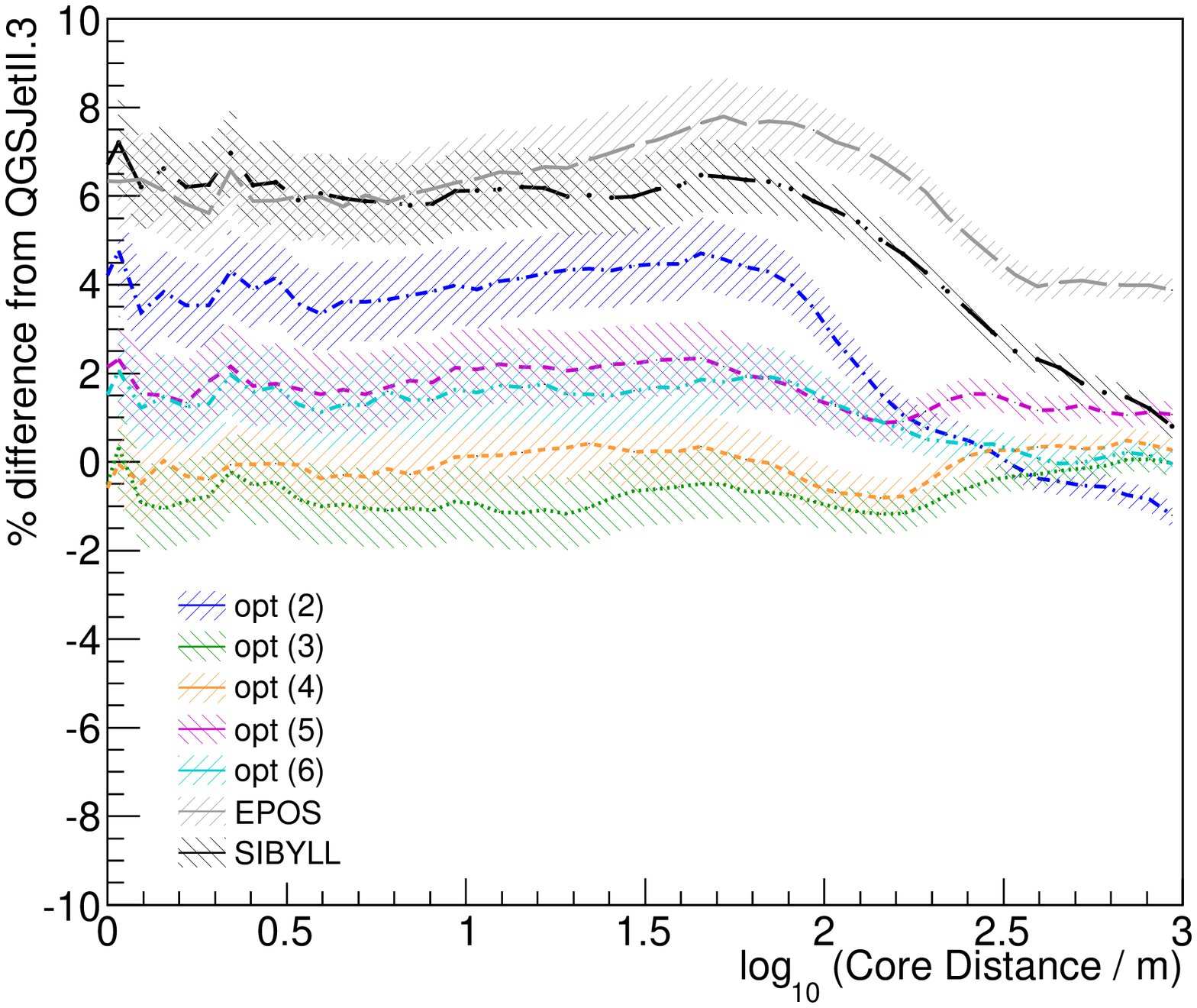} \hfill
\includegraphics[width=0.49\textwidth]{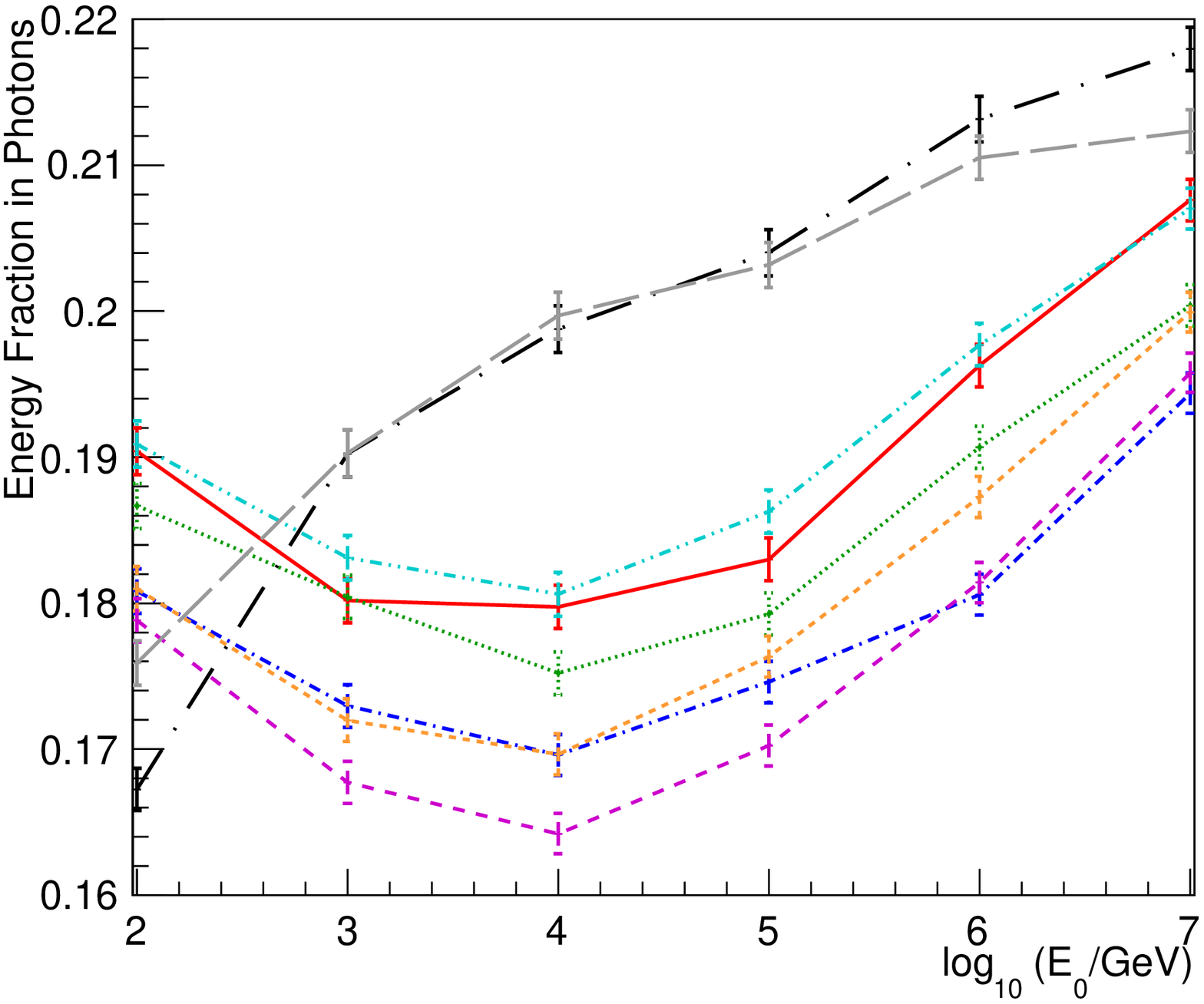}
\end{center}
\vspace*{-0.5cm}
  \caption{\textbf{Left:}  Percentage
  difference of the Cherenkov photon
  density $\rho$ (for $\lambda$ = 300-600 nm) compared to the standard QGSJET-II model
  (i.e. $\rho_{\rm i}/\rho_{\rm std}) - 1$) at 2000 m altitude. Results from $10^{4}$
  vertical proton showers of $10^{12}$ eV. Atmospheric absorption is accounted for. 
  Error bars show statistical errors of the mean values.
 \textbf{Right:} Fraction of total interaction energy emitted in photons 
  as a function of energy ($E_{\gamma}/E_{\rm tot}$)}
\label{fig-cherplots}
\end{figure}

However, as the number of shower particles reaching ground level at this energy
is very low, a more appropriate observable is
the lateral distribution of Cherenkov photons at ground level. The number of Cherenkov 
photons relates to the primary energy, but is different for different
primaries. Proton induced air showers are the main background in
gamma ray experiments using the atmospheric Cherenkov technique and 
their simulated appearance depends on the hadronic interaction models.
Small differences are seen in the lateral distributions (Fig.~\ref{fig-cherplots}), 
resulting from the differing fractions of
energy that is channeled into the electromagnetic (EM) component. 
This is visible in QGSJET option 2, where 
the biggest fraction of energy is going into the EM component, 
leading to the largest number of
Cherenkov photons and lowest number of muons.
SIBYLL and EPOS predict the largest numbers of Cherenkov photons, with 
about 7\% more than the standard
QGSJET option.  This difference alone contributes 7\% to the systematic energy
error of proton showers in Cherenkov telescope experiments.
The systematic shift is lower
than the energy resolution of current instruments ($\approx$ 15\% for 1 TeV photons),
so it should not affect much measurements of the background
 in ACT instruments.

The ability to identify electromagnetic gamma ray showers from 
the overwhelming background of hadronic events 
directly determines the sensitivity of a Cherenkov telescope.
It is based on the analysis of the shape of the shower image
and depends on hadronic models, too.
Hadronic showers produce more diffuse and larger images in 
the camera of a Cherenkov telescope than compact electromagnetic showers.
Those hadron showers that are misidentified as gamma showers 
are usually characterised by a large 
fraction of the shower energy being dumped early on into the EM component \cite{EMfrac}. 


The EM fraction in Fig. \ref{fig-cherplots}
looks similar to the inelasticity shown in  Fig. \ref{fig-cross},
with values of about 1/3 of the ones of the inelasticity,
as one would expect if 1/3 of the pions produced are $\pi^0$s.
Again, most QGSJET variants produce a lower EM fraction than 
the standard QGSJET model,
 and SIBYLL and EPOS show different shapes.
Overall, SIBYLL and EPOS were found to produce a larger EM fraction than QGSJET
above $10^{12} $ eV. 
A larger EM fraction in the first interaction 
means a more photon-like appearance of the shower and 
a higher chance to misidentify a proton-induced shower as a gamma ray. 
Thus, all QGSJET variants give an equal or smaller rate of photon-like events
than the QGSJET standard. 
Both, SIBYLL and EPOS will produce more gamma-like 
proton showers, but to quantify this would require full simulation of 
the telescope array and the reconstruction procedure.

This difference in the EM fraction can be partly explained by the
differences in the inelasticity of the models, changing the amount of 
energy available for particle production. 
Once this effect was accounted for the QGSJET variants were found to give here a constant value of about 31\%
whereas SIBYLL is about 1\% higher and EPOS 1\%  lower.
As EPOS leaves more energy in secondary baryons, it is not 
surprising that the fraction in $\pi^0$s, and thus photons, 
is smaller.

\subsection{Medium Energies ($10^{15}$ eV)}

\begin{figure}[t]
\begin{center}
\includegraphics[width=0.49\textwidth]{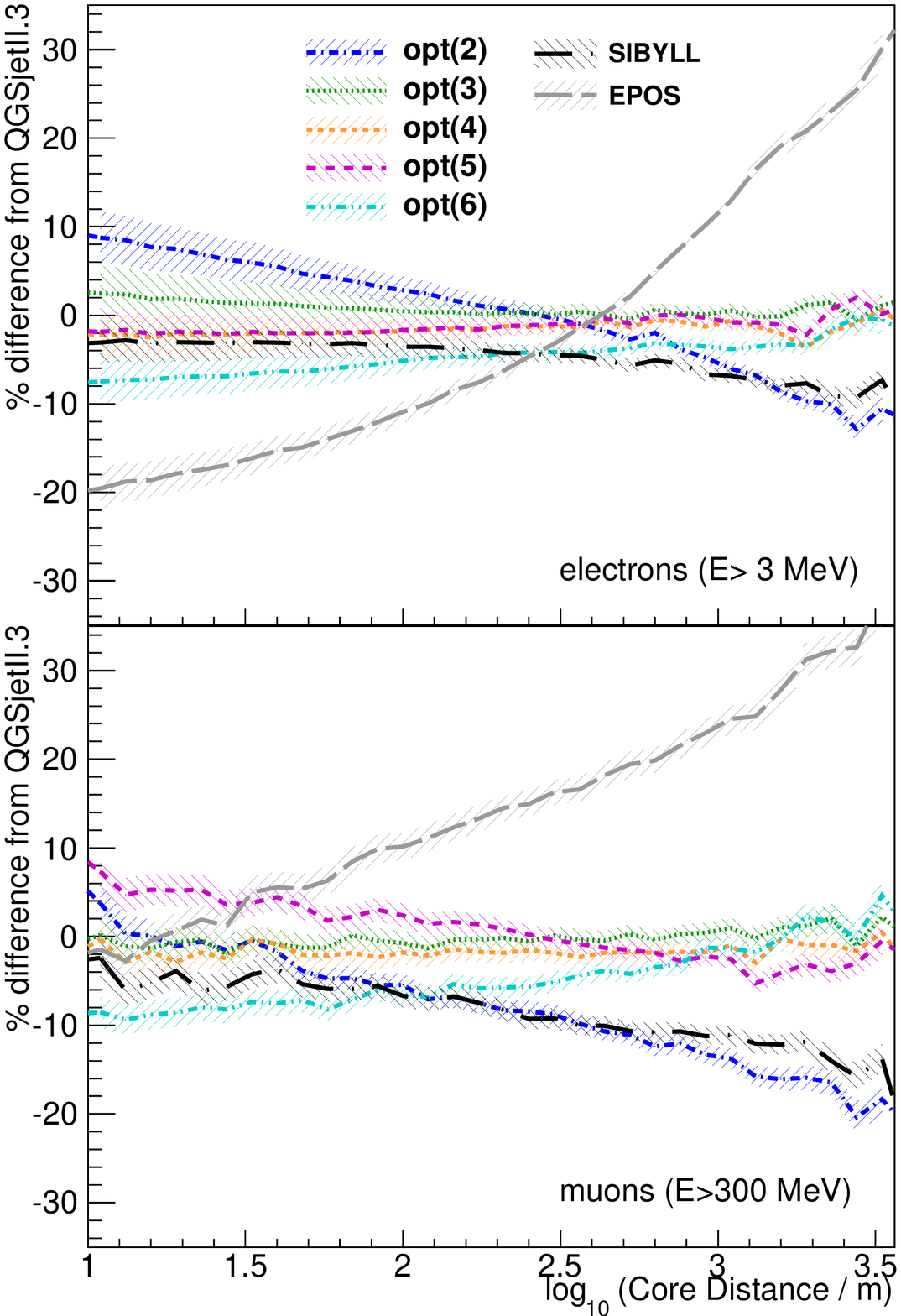} 
\includegraphics[width=0.49\textwidth]{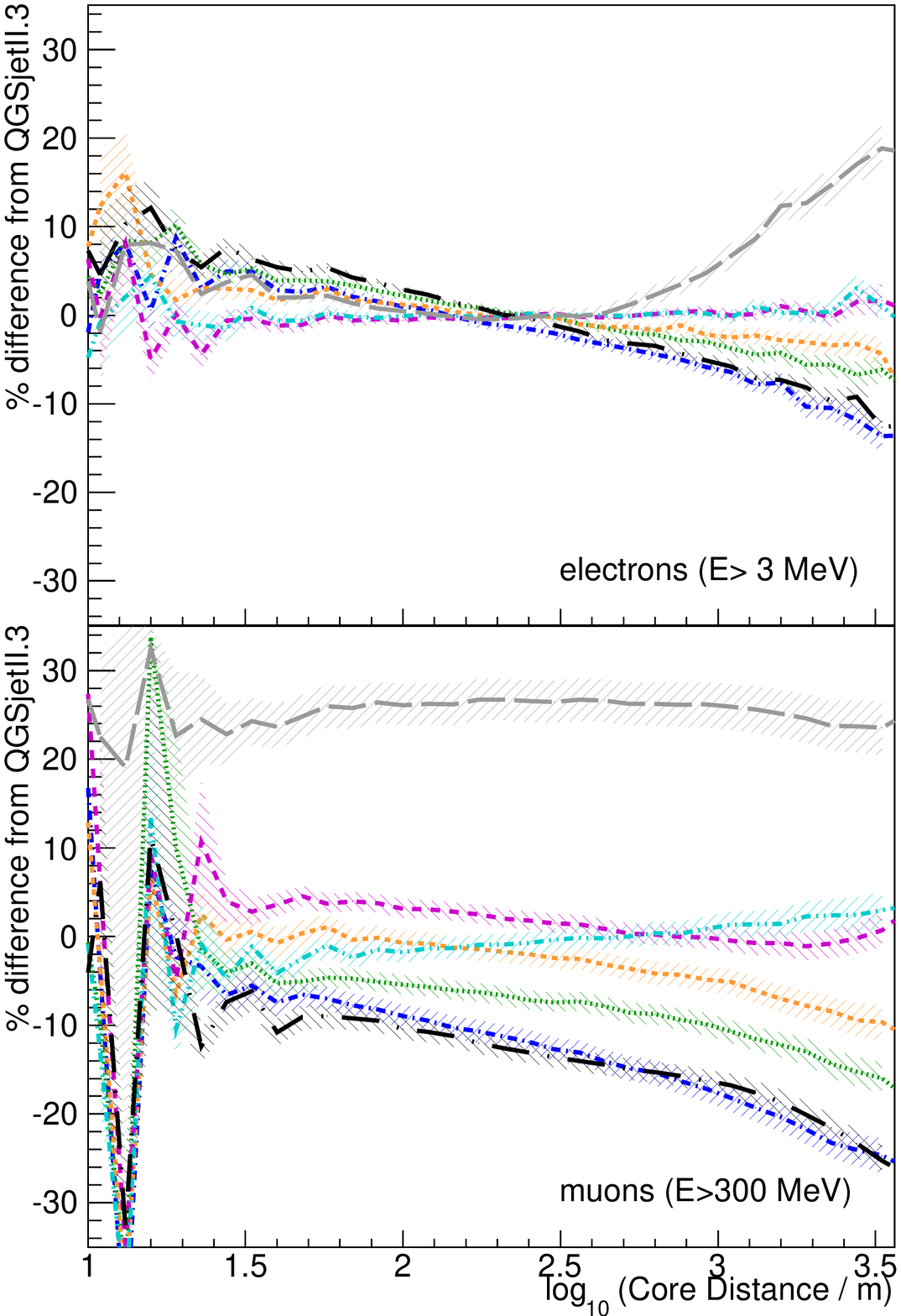} \\
\includegraphics[width=0.49\textwidth]{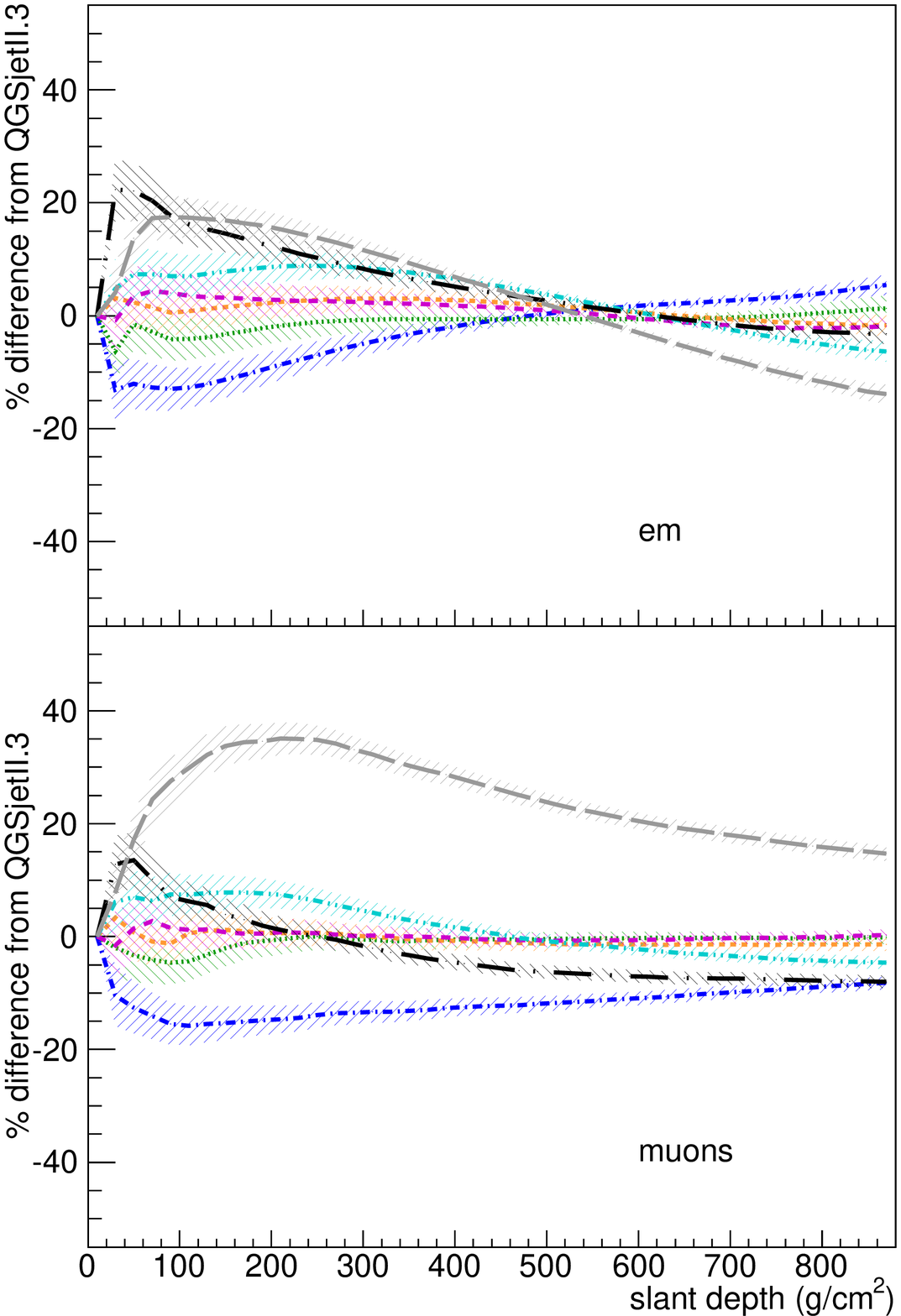} 
\includegraphics[width=0.49\textwidth]{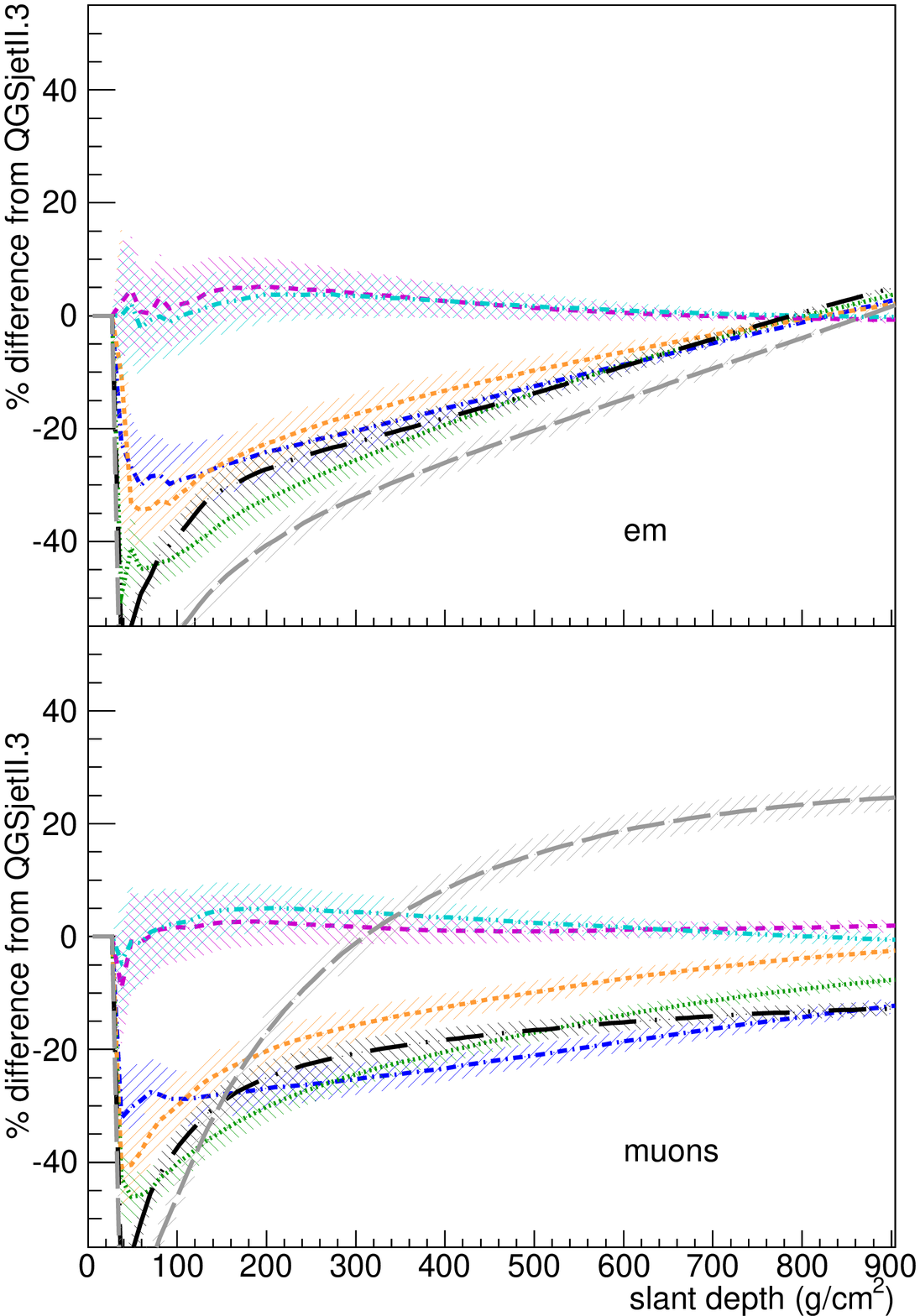}
\end{center}
\vspace*{-0.5cm}
  \caption{\textbf{Top:} Average percentage difference (compared to standard
  QGSJET-II) of the
  particle number $N$ as function of the core distance at 1452 m altitude
  (i.e. $N_i/N_{\rm std} - 1$) for electrons and muons. 
  \textbf{Bottom:} 
  Average percentage difference (compared to standard QGSJET-II) of the
  ionisation energy deposit as a function of the atmospheric depth 
  for the em component and for muons.  
  \textbf{Left:}  300 vertical proton showers of $10^{15}$ eV. 
  \textbf{Right:} 100 proton showers at $20^{\circ}$ with $10^{19}$ eV. 
  Shaded error bands show  statistical errors of the mean values.}
\label{fig-showerplots}
\end{figure}

\begin{table}[t]
\begin{center}
\begin{tabular}{|l|c|c|c|c|c|}
\hline 
Model & \rule{0mm}{4mm}
$N_e(10^{3})$ & $N_\mu(10^{3})$ & $N_e/N_\mu$ &  $X_{\rm max}$ (g/cm$^2$) & $N_{\rm max}(10^{3})$ \\  
\hline 
\hline 
QGSJET-II~(1)& $208 \pm4$ & $10.3 \pm 0.1$ & $20.8 \pm 0.4$ & $546 \pm 3$  & $ 566 \pm 4$ \\
QGSJET-II~(2)& $219 \pm 4$ & $9.4 \pm 0.1$ & $24.0 \pm 0.5$ & $552 \pm 4$  & $ 577 \pm 4$ \\
QGSJET-II~(3)& $211 \pm 5$ & $10.3 \pm 0.1$ & $21.6 \pm 0.8$ & $548 \pm 4$  & $ 571 \pm 3$ \\ 
QGSJET-II~(4)& $203 \pm 4$ &  $10.1 \pm 0.1$ & $21.0 \pm 0.5$ & $542 \pm 3$  & $ 574 \pm 3$ \\ 
QGSJET-II~(5)& $204 \pm 4$ & $10.3\pm 0.1$ &  $20.1 \pm 0.4$ & $542 \pm 3$  & $ 567 \pm 3$ \\ 
QGSJET-II~(6)& $195 \pm 4$ & $9.8\pm 0.1$ & $20.6 \pm 0.5$ & $534 \pm 3$  & $ 579 \pm 3$ \\
\hline
SIBYLL~2.1  &  $201 \pm 4$ & $9.4 \pm0.1$ & $22.1 \pm 0.5$ & $539 \pm 3$  & $ 571 \pm 4$ \\ 
EPOS~1.99  & $178 \pm 3$ & $11.8 \pm 0.1$ & $16.3 \pm 0.6$ & $528 \pm 3$  & $ 564 \pm 4$ \\ 
\hline
\end{tabular}
\end{center}
 \vspace*{-0.5cm}
 \caption{\rm Simulated mean observables for    
  proton showers of energy 10$^{15}$ eV at 0$^\circ$.
  The errors given are the  statistical errors of the mean values.
  Note that the position of the first interaction was fixed at a depth of 33 g/cm$^2$.}
\label{tab-10^15obs}
\end{table}

At medium energies ($\sim 10^{15}$ eV) detector arrays usually measure
shower particles arriving at ground level, with some arrays being able
to distinguish between e$^{\pm}$, photons and muons.
Therefore, Tab.~\ref{tab-10^15obs} gives some typical shower observables
for the different models. 
The left side of 
Fig.~\ref{fig-showerplots} shows differences in electron and muon numbers
as function of core distance (lateral distributions) and ionisation
energy deposit as a function of atmospheric depth (longitudinal
distributions) for $10^{15}$ eV proton showers.
At $10^{15}$ eV differences between the QGSJET model variations are less than
$\pm$10\% for electrons and muons at core distances of 10 to 1000 m
(the most relevant core distances at these energies are $\le$ 200 m).  
The longitudinal distributions show very good agreement around the
shower maximum ($\sim$550 g/cm$^2$) and differences up to $\pm10$\% at
ground level  (here, 880 g/cm$^2$).
  SIBYLL lies well within the spread of the QGSJET
options, but EPOS shows  much flatter lateral distributions, 
for both electrons and muons, with many more particles at large   
distances from the shower core and overall about 20\% more muons. 
At ground level EPOS has 15\% fewer electrons and 15\% more
muons than the reference model. 
Consequently, all energy assignments would be different on the 15-20\% level
and composition analysis were greatly affected by these differences.
They are big enough for the KASCADE experiment to tell the difference.
Indeed, there is evidence from KASCADE data  \cite{haungs_arena}
that the primary proton spectrum
agrees better with direct measurements
when QGSJET and SIBYLL are used for the analysis, than when EPOS is used.

The lateral distribution of QGSJET option 2 (enhanced diffraction) is very
similar to that of SIBYLL at both $10^{15}$ and $10^{19}$ eV, 
suggesting that one of the major differences between the two models is 
the rate of diffraction. However
differences can be seen between option 2 and SIBYLL in the results of
single particle interactions, indicating that there are other important 
physical  distinctions.

\subsection{Ultra High Energies ($10^{19}$ eV)}

\begin{table}[t]
\begin{center}
\begin{tabular}{|l|c|c|c|c|}
\hline 
Model \rule{0mm}{4mm}& $S_{1000}$ (GeV) & $t_{1/2}$ (ns) & $X_{\rm max}$ (g/cm$^2$) & $N_{\rm max}(10^{7})$ \\  
\hline 
\hline 
QGSJET-II~(1)&  $10.01 \pm 0.09$ & $373.2 \pm 1.4$ & $773 \pm 6$ & $647 \pm 2$\\
QGSJET-II~(2)&  $ 8.86 \pm 0.11$ & $385.7 \pm 1.7$ & $801 \pm 6$ & $642 \pm 2$\\
QGSJET-II~(3)&  $ 9.38 \pm 0.09$ & $382.2 \pm 1.4$ & $803 \pm 6$ & $645 \pm 2$\\ 
QGSJET-II~(4)&  $ 9.68 \pm 0.10$ & $378.9 \pm 1.4$ & $794 \pm 5$ & $645 \pm 2$\\ 
QGSJET-II~(5)&  $10.01 \pm 0.08$ & $371.6 \pm 1.4$ & $780 \pm 6$ & $644 \pm 2$\\ 
QGSJET-II~(6)&  $10.07 \pm 0.10$ & $372.9 \pm 1.3$ & $784 \pm 5$ & $648 \pm 2$\\
\hline
SIBYLL~2.1   &  $ 9.00 \pm 0.09$ & $388.0 \pm 1.7$ & $795 \pm 5$ & $651 \pm 2$\\ 
EPOS~1.99    &  $11.61 \pm 0.18$ & $368.9 \pm 2.0$ & $804 \pm 5$ & $630 \pm 2$\\ 

\hline
\end{tabular}
\end{center}
\vspace*{-0.5cm}
  \caption{Simulated mean observables for    
  proton showers of energy $10^{19}$ eV at $20^{\circ}$. 
  The errors are the  statistical errors of the mean values.
  Note that the position of the first interaction was fixed at a depth of 38 g/cm$^2$.}
\label{tab-10^19obs}
\end{table}
At $10^{19}$ eV the detector arrays must be sparse and no longer the 
total particle number is measured, but the particle density as a function
 of 
distance from the core. The energy is then determined from the density at some
distance \cite{newton}. The time structure of the shower front, measured in the 
array detectors, carries information on the shower development and the primary mass.
With fluorescence telescopes the longitudinal profile and the position 
of the shower maximum can be directly measured, from which the energy and
the composition can be inferred. 
Commonly used observables at 10$^{19}$ eV are listed
in Tab. \ref{tab-10^19obs}.

As expected, the deviations from standard QGSJET are largest at this energy, 
as the extrapolations of model parameters from collider energies are
largest. Deviations in the lateral distributions of electrons
and photons are less than +10\% within 1000 m from the shower axis.
For muons the variations within the QGSJET models are less than 20\%
at 1 km core distance, but all the options produce always smaller muon
numbers than the standard QGSJET. SIBYLL gives almost the same
results for electron and muon lateral distributions as QGSJET option
2, the version with the enhanced proton diffraction. Again, EPOS is
distinctly different, with 25\% more muons at ground level for all core
distances.  The QGSJET and SIBYLL muon numbers at ground are within 0
to -15\% of the standard version.  The EPOS electron numbers at
ground are similar to those of the other models, owing to the fact that
for the angle chosen the shower maximum is close to the ground.
For more inclined showers larger differences would occur.

In the longitudinal distributions, the QGSJET options 2, 3 and 4 and SIBYLL
have markedly lower particle numbers and energy deposits (40-20\%) in
the upper part of the shower than the standard.  EPOS shows a similar trend,
again developing slowly early in the shower, but quickly producing larger
numbers of muons, with almost 30\% more muons at ground level.

To reproduce data from experiments directly measuring
muon numbers such as KASCADE and HiRes-MIA and from the Pierre Auger Observatory,
where the muon content can be inferred indirectly
\cite{universality},
more muons in simulations seem to be 
required. EPOS seems to address the muon deficit problem, 
but would introduce a significantly different energy scale.

Significant differences are also seen in $S_{1000}$, a common
observable used for the energy reconstruction in the $10^{19}$ eV region
(see Tab. \ref{tab-10^19obs}).
The energy reconstruction based on option 2
would give $\sim$10\% higher cosmic ray energies than the one using the
standard QGSJET option, while EPOS would give 16\% smaller energies.
The differences in average $X_{\rm max}$ for the QGSJET models are
less than 30 g/cm$^2$ (4\%), with all the models producing deeper 
$X_{\rm max}$ values than standard QGSJET. 
The differences in $t_{1/2}$, which is used for composition analysis, 
are less than 5\%, with SIBYLL and EPOS giving the extreme values.

\section{Conclusion}

The analysis of shower simulations demonstrates the increase in systematic uncertainties 
(introduced by variation of model parameters) as the primary particle energy increases,
and with it the extrapolation of the models.
Our study indicates that parameter variations within one model 
generally cannot reproduce the full variance possible between models.
None of the parameter variations investigated in QGSJET was able to 
mimic the behaviour of the EPOS model: 
At any energy, EPOS has far more muons than all other models
and specifically than the QGSJET standard version.
Differences in several observables amount to 
$<$ 10\% in the TeV range and to 20-30\% in the PeV and EeV ranges, causing sizeable 
systematic errors on the energy scale and composition analysis.
Within QGSJET, at all primary energies the largest effects are observed
due to the increase of the rate of diffraction for protons and pions,
making diffraction the feature with the largest impact on the
overall development of the shower. 

EPOS is a relatively new model and still has to prove its qualities
by being compared in detail to results from different (past and present) 
experiments from TeV to EeV energies.

The lateral and longitudinal distribution pattern of 
the electromagnetic and muonic components in air showers
shows complex variations between models, which depend on the energy
and the level of shower development, the zenith angle, etc.
Therefore, complete simulations of showers, detectors and analysis procedures,
using different interaction models  
are needed for a meaningful data analysis.
The model uncertainties at high energies are of a size that makes 
analysis of nuclear composition at $\geq$EeV energies rather difficult 
(with the observables considered here), 
as differences between proton and iron induced showers are not much larger 
than the uncertainties due to the models.
On the other hand, the uncertainties are large enough that 
current experiments are able to tell which model fits best.
Also, new data from LHC and RHIC do increasingly constrain the models, 
and help to make the extrapolation to highest energies more reliable.
But ultimately, only a coherent description of 
cosmic ray phenomena and hadronic interaction
physics over the full energy range
will be a convincing proof of the correct interpretation of cosmic ray data.

\section*{Acknowledgments}

We gratefully acknowledge the support of: The Science and Technology
 Facilities Council, the Henry Ellison Scholarship Fund and the
 program Romforskning of Norsk Forsknigsradet.

\section*{References}


\begin{thebibliography}{99}
   
 \bibitem{GRtheory}  V.N.~Gribov,  Sov.~Phys.~JETP~26 (1968) 414

 \bibitem{Glauber} R.J.~Glauber, G.~Matthiae, 
 Nucl.~Phys.~B 21 (1970) 135

 \bibitem{GRnucl}  V.N.~Gribov,  Sov2.~Phys.~JETP~29 (1969) 483

 \bibitem{ulrich2010} R.~Ulrich, R.~Engel and M.~Unger,  arXiv:1010.4310 [hep-ph] 
 
 \bibitem{qgsjet}
    S.~Ostapchenko, Phys.~Rev.~D 74 (2006) 014026 and \\
    AIP Conf.~Proc.~928 (2007) 118

  \bibitem{sibyll} R.S.~Fletcher et al., Phys.~Rev.~D 50 (1994) 5710\\
   E.J.~Ahn et al., Phys.~Rev.~D 80 (2009) 094003

  \bibitem{epos} K.~Werner et al., 
  Phys.~Rev.~C 74 (2006) 044902

  \bibitem{hla01} M.~Hladik et al., 
  Phys.~Rev.~Lett.~86 (2001)  3506 

  \bibitem{pierog2008} T.~Pierog and K.~Werner,
  Phys.~Rev.~Lett.~101 (2008)   171101 

  \bibitem{corsika} D.~Heck et al., 
  Report FZKA 6019 (1998), Forschungszentrum Karlsruhe
  
 \bibitem{goo60} 
M.L.~Good and W.D.~Walker,   Phys.~Rev.~120 (1960) 1857 

\bibitem{kai79} 
 A.B.~Kaidalov, Phys.~Rep.~50 (1979) 157
 
\bibitem{kha95} D.~Kharzeev,  Phys.~Lett.~B~378 (1996) 238 

  \bibitem{david2011} D.~d'Enterria  et al., arXiv:1101.5596v2 [astro-ph.HE]

 \bibitem{lhcf}
    A.~Tricomi for the LHCf Collaboration, Talk at 35$^{\rm th}$ ICHEP, 
    Paris (2010) \\
    Early Physics with the LHCf detector at LHC\\
    http://indico.cern.ch/contributionDisplay.py?contribId=860\&confId=73513

  \bibitem{HESSexp} J.A.~Hinton, 
  New Astronomy Review 48  (2004) 331

  \bibitem{VERITASexp} T.C.~Weekes et al., 
  Astrop.~Phys.~17 (2002) 221

  \bibitem{MAGICexp} E.~Lorenz, 
  New Astronomy Review 48 (2004) 339
  
  \bibitem{KASCADEexp}  T.~Antoni et al. (KASCADE Collaboration), 
   Nucl. Instr. Meth. A 513 (2003) 490

   \bibitem{lhaaso}
   Z.~Cao et al.~(LHAASO Collaboration), 
   Proc.~31$^{\rm st}$ ICRC (2009), \L{}\'od\'z, Poland \\
   http://icrc2009.uni.lodz.pl/

  \bibitem{ta}
   J.N.~Matthews et al.~(Telescope Array Collaboration), 
   Proc.~31$^{\rm st}$ ICRC (2009), \L{}\'od\'z, Poland \\
   http://icrc2009.uni.lodz.pl/
   
  \bibitem{Augerexp}  J.~Abraham et al.~(Pierre Auger Collaboration), 
  Nucl.~Instr.~Meth.~A 523 (2004) 50
  
  \bibitem{s1000} J.~Abraham et al.~(Pierre Auger Collaboration), 
  Phys.~Rev.~Lett.~101 (2008) 061101 
  
  \bibitem{RiseTime} J.~Abraham et al.~(Pierre Auger Collaboration), 
  Astrop.~Phys.~29 (2008) 243-256 
    
  \bibitem{EMfrac} G.~Maier, J.~Knapp,  
  Astrop.~Phys.~28 (2007) 72
  
  \bibitem{haungs_arena} A.~Haungs for the KASCADE-Grande Collaboration, 
  Talk at ARENA Meeting, Nantes (2010) \\
  Latest results and perspectives of the KASCADE-Grande EAS facility\\
  http://indico.in2p3.fr/contributionDisplay.py?contribId=17\&confId=2719
  
  \bibitem{newton} 
  D.~Newton et al.,  Astrop.~Phys.~26 (2007) 414
  
  \bibitem{universality} F.~Schmidt et al., 
  Astrop.~Phys.~29 (2008) 355

  
  
\end{thebibliography}
\end{document}